\documentclass[11pt,twoside]{article}
\usepackage{./asp2010}
\usepackage{graphicx}

\resetcounters

\begin{document}

\title{The IMF of Field OB Stars in the Small Magellanic Cloud}
\author{J. B. Lamb, M. S. Oey, A. S. Graus, and D. M. Segura-Cox
\affil{University of Michigan \\ 
Department of Astronomy, 830 Dennison Bldg., Ann Arbor, MI 48109-1042, USA}}

\begin{abstract}
The population of field OB stars are an important component of a galaxy's stellar content, representing 20-30\% of the massive stars. To study this population, we have undertaken the Runaways and Isolated O Type Star Spectroscopic Survey of the SMC (RIOTS4).  RIOTS4 surveys a spatially complete sample of $>$350 field OB stars in the Small Magellanic Cloud and will serve as a key probe of runaways, binaries, and the stellar IMF in the field massive star population.  Here, we focus on the field IMF, which provides an empirical probe of the star-forming process and is a fundamental property of a stellar population.  Together with photometry from the OGLE survey, RIOTS4 will yield a definitive stellar IMF for the SMC field massive star population.  We present preliminary results that suggests the field IMF is much steeper, $\Gamma$ = 2.9, than the canonical stellar IMF of $\Gamma$ = 1.35.  Despite the steep slope, we see no evidence of a stellar upper mass limit, up to our highest mass star of $65M_\odot$.    
\end{abstract}

\section{Introduction}
From an
observational standpoint, our understanding of massive star formation is mostly derived from the
products of this process: the stars themselves.  One of the 
fundamental observable properties of star formation is the
stellar initial mass function (IMF).  For high mass stars, the IMF
follows a simple power law given by ${dn} / {d\log m} \propto m^{-\Gamma}$
where $n$ is number of stars, $m$ is stellar mass and $\Gamma$ = 1.35 (Salpeter,
1955).  This Salpeter IMF is uniformly applicable to a large variety
of cluster environments, from OB associations to super star clusters.
However, the field massive stars are one astrophysical
environment where the IMF slope may deviate from Salpeter.
Despite the robust
nature of the Salpeter slope in clusters, there is evidence that the
IMF of field massive stars is much steeper than Salpeter, with
measurements of $\Gamma$ = 3 to 4 in the Magellanic Clouds (Massey et
al. 1995, Massey 2002).  Within the Galaxy, van den Bergh (2004) confirms that the
field population is biased towards later spectral types, indicating they are
either older or less massive than the cluster population.  However, the field population around the 30 Dor region of
the Large Magellanic Cloud exhibits a Salpeter slope (Selman et al. 2011), 
and $\Gamma = 1.8$ in NGC 4214 (Ubeda et al. 2006).
Thus, the null hypothesis of a universal IMF slope for both clusters
and the field cannot easily be rejected. 

We believe that the field massive stars consist of both in-situ stars
and runaway stars formed in clusters (Oey \& Lamb, this volume).
Thus, the IMF parameters, which include both the slope and the
upper-mass limit, are important diagnostics for theories of massive 
star formation.  For example, competitive accretion models
require that massive stars only form in clusters, and that the mass of
the most massive star in a cluster $m_{\rm max}$ is related to
the total cluster mass $m_{\rm cl}$ by $m_{\rm max} \propto m_{\rm cl}^{2/3}$
 (e.g. Bonnell et al. 2004).  Whereas, monolithic collapse
models more readily allow for the formation of isolated massive star
formation is (e.g. Krumholz \& McKee 2008).  

To investigate the mass function of field massive stars, we
present here first results from the Runaways and Isolated O Type Star
Spectroscopic Survey of the SMC (RIOTS4; Oey \& Lamb, this volume).
In this contribution, we present the slope 
and upper mass limit of the stellar IMF for field massive stars, which
may provide key insights about the process of massive star formation
and illuminate any differences between isolated and clustered star
formation.   

\section{RIOTS4 Observations}

RIOTS4 is an extensive spectroscopic survey of a spatially complete
sample of field massive stars, which covers the entire star-forming
area of the Small Magellanic Cloud (SMC).  The initial sample of all
massive stars in the SMC was selected using the reddening-free
parameter $Q$ $\le -0.84$ and $B$ $\le$ 15.21 from the $UBVR$
photometry of Massey (2002).  To identify the field stars, Oey et
al. (2004) ran a friends-of-friends algorithm (Battinelli, 1991) on
the OB candidate sample, which sets a physical clustering length such that
the total number of clusters, defined as $n_{\rm stars} \ge 3$,  is
maximized.  The field population are those OB stars that are at least
one clustering length (28 pc) away from the nearest OB star.  We have
obtained spectra for the majority of these stars with the IMACS
multi-object spectrograph ($R \sim$ 2600) on the Magellan-Baade
telescope.  Our completeness is nearly 100\% for the SMC bar and over
85\% for the SMC wing.   

We obtain spectral types from our spectra following the classification scheme and atlas prepared by Walborn \& Fitzpatrick (1990).  
Spectral types are converted into stellar effective temperature, $T_{\rm eff}$, using the calibration for SMC O stars from Massey et al. (2005) and for B stars from Crowther et al. (1997).  The Crowther conversion was chosen due to the smooth transition at the overlapping B0 spectral type with the Massey calibration.  
The absolute $V$ magnitudes of our stars are obtained from the
photometry of Massey 
(2002), assuming a distance modulus of 18.9 (Szewczyk et
al. 2009) and applying SMC extinction maps from Zaritsky et al. (2002).   
We convert the absolute $V$ magnitude into a bolometric magnitude, $M_{\rm bol}$, using $T_{\rm eff}$ to calculate the bolometric correction following the equations of Massey et al. (2005).  

\begin{figure}
	\begin{center}
	\includegraphics[scale=.44,angle=0]{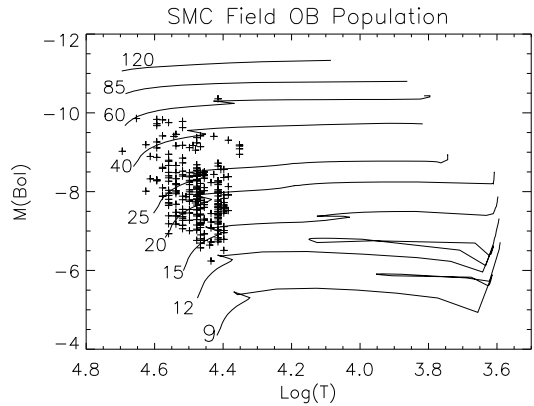}
	\caption{An H-R diagram of our field stars with physical parameters derived from spectral types and Massey (2002) photometry.  The evolutionary tracks plotted here are from Charbonnel et al. (1993) at SMC metallicity and are labelled corresponding to the stellar mass of the evolutionary track in solar masses.}
	\label{photcmd}
	\end{center}
\end{figure}

Using the derived $T_{\rm eff}$ and $M_{\rm bol}$, we plot an H-R
diagram of our field massive stars in Figure 1.  Additionally, we
plot Geneva stellar evolutionary tracks at SMC metallicity ($Z = 0.004$;
Charbonnel et al. 1993), which are labelled according to the stellar
mass of the evolutionary track.  Our survey is complete to $25M_\odot$ along the
Zero Age Main Sequence (ZAMS).
One notable feature of this H-R
diagram is the offset of our observations from the theoretical main
sequence provided by the evolutionary tracks.  This problem is not
unique to our data, however, as spectroscopic data for massive SMC
stars plotted in Massey (2002) are similarly plagued by this issue.
It is unclear whether this is an issue with the observational data, or a
problem with the SMC metallicity stellar evolutionary models.

\section{Field Massive Star IMF}
To construct the IMF of the field population, we follow the formalism of Koen (2006), where the IMF is constructed as a cumulative distribution function (CDF) given by
\begin{equation}
F(m) = \int_L^U m^{-(\Gamma+1)}\, \mathrm{d}m =  \frac {j} {(N+1)}
\end{equation}
with $L$ and $U$ being the lower and upper stellar mass limits.  The
final term represents an empirical CDF using a ranked order of stellar
masses, $j$, where $j = 1$ is the lowest mass star, $j = N$ is the
highest mass star, and $N$ is the number of stars in our sample.  From
this empirical CDF, we can reconstruct a mass function by plotting
$\log [1 - F(m)]$ versus $\log m$, which is shown in Figure 2.
However, since we are dealing with field stars, we are actually
measuring the present day mass function (PDMF) rather than the IMF.
We measure the slope of the PDMF using two methods:  a linear least
squares fit and a maximum likelihood method given by Koen (2006; his
equation 10).  The least squares fit yields a PDMF slope of $\Gamma$ =
3.8 while the maximum likelihood method yields $\Gamma$ = 3.2.  We
note that the PDMF is linear across the entire mass range, from our
completeness limit of $25M_\odot$ to our most massive star at
$65M_\odot$.  The absence of a high mass turn-off in this distribution
indicates that we do not observe an upper mass limit in the field.
This is consistent with the universal upper-mass limit found in Milky
Way and LMC clusters, $\sim 150M_\odot$ (Oey \& Clarke 2005; Koen 2006).

\begin{figure}
	\begin{center}
	\includegraphics[scale=.44,angle=0]{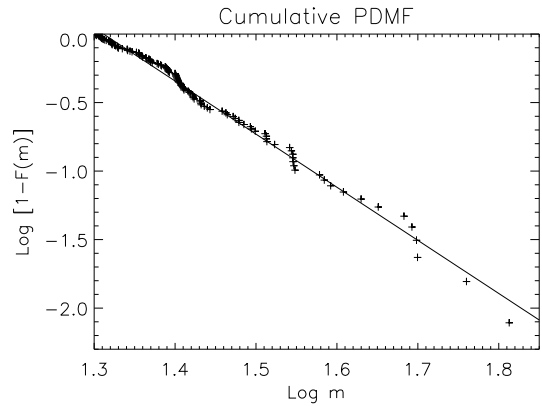}
	\caption{The PDMF of the SMC field star population, plotted as $\log[1-F(m)]$ versus $\log m$, where $F(m)$ is the empirical CDF.  The plotted line shows the linear least squares fit to the data.}
	\label{koen}
	\end{center}
\end{figure}

To derive the intrinsic field IMF, we devise a simple Monte Carlo code
to generate theoretical stellar populations with the assumption of
continuous star formation and an IMF slope varied from $\Gamma_{\rm
  IMF}$ = 2.0 to 4.0 in steps of 0.1 and $10^4$ iterations at each
step.  For each input IMF slope, we compare the distribution of PDMF
slopes from the model to the observed PDMF slope, finding a best fit
when the input $\Gamma_{\rm
  IMF}$ = 2.9.  Although this slope is significantly steeper than the
canonical Salpeter IMF, it is in agreement with the field IMF found in
the Magellanic Clouds by Massey et al. (1995) and Massey (2002).   

\section{ IMF for $7 - 20\ M_\odot$ Stars}
We plan to supplement our high mass field IMF by measuring the field IMF from $7-20 M_\odot$ using photometric data from the Optical Gravitational Lensing Experiment (OGLE; Udalski et al. 1998, 2008).  OGLE provides BVI photometry of the SMC bar and VI photometry for the entire SMC, with completeness to $7M_\odot$, allowing for up to two magnitudes of extinction.  However, it is difficult to extract individual stellar masses from just BVI photometry, so we will adopt a statistical approach to determine the IMF.  We create $10^4$ realizations of each star in the OGLE database, which are given random gaussian photometry and extinction errors, as defined by the observational uncertainties.  Using the stellar evolution tracks of Charbonnel et al. (1993) and their conversion into optical colors by Girardi et al. (2002), we can count the number of realizations for each star that falls into different mass bins and in this manner, assign each star a fractional probability of belonging to specific mass bins.  

OGLE also will allow us to redefine our field star sample using a more
stringent mass constraint for isolation.  Instead of requiring a field OB star to be at least one clustering length from other OB stars, we can strengthen the requirement to at least one clustering length from stars  $>7M_\odot$. This analysis will allow us to examine how the method for defining a field star affects the IMF of the field
population. 

\section{Discussion and Conclusions}
We measure the IMF of our spatially complete sample of field massive
stars in the SMC to be $\Gamma$ = 2.9, much steeper than the canonical
Salpeter IMF of $\Gamma$ = 1.35.  We find no evidence of an upper
stellar mass limit for field stars, up to our most massive star of
$65M_\odot$.  This suggests that despite the steep IMF, the field is
not limiting the formation of the most massive stars.
Our results confirm the
steep field IMF that Massey et al. (1995) and Massey (2002) found in
the both the Magellanic Clouds.  This steep IMF may indicate a
different mode of star formation is happening in the field.

However, our results are affected by a number of uncertainties.  Firstly, there is
some concern in the ability of evolutionary models to properly
estimate stellar masses (Massey et al. 2005). Secondly, the contribution from
runaway stars in both their number and mass function is also an
unknown quantity.  However, since the fraction of O stars that are runaways
is higher than B stars (Gies \& Bolton 1986), the runaway population 
tends to flatten, rather than steepen, the IMF.  
Thus, runaways will affect, perhaps significantly, our observed IMF
compared to the IMF of in situ field stars, which may be even steeper than our measurements.
  Other sources of uncertainty arise from Oe/Be stars in our survey, whose stellar
emission lines make spectral classification difficult or
impossible; and finally, the binary fraction of the field, which the
preliminary results from our survey suggest is $>$ 50\% (Oey \&
Lamb, this volume), may also affect the IMF.  Future results from our survey will better
quantify the degree and magnitude of the uncertainty that runaways,
binaries, and Oe/Be stars impart on the stellar IMF of the field.

\nocite{2005ApJ...627..477M,2002ApJS..141...81M,1995ApJ...438..188M,1955ApJ...121..161S,2007AJ....133..932U,2004AJ....128.1880V,2011ASPC..440...39S,2004MNRAS.349..735B,2008Natur.451.1082K,2004AJ....127.1632O,1991A&A...244...69B,1990PASP..102..379W,1997IAUS..189..137C,2009AJ....138.1661S,2002AJ....123..855Z,1993A&AS..101..415C,2006MNRAS.365..590K,1998AcA....48..147U,2008AcA....58..329U,2002A&A...391..195G,2003ARA&A..41...57L,1986ApJS...61..419G,2005ApJ...620L..43O}

\acknowledgements 
I thank the organizers  for a delightful venue and quality science program.
This work was supported by funding from NSF grant AST-0907758.  Travel support was provided by the University of Michigan Rackham Graduate School.

\bibliography{Lamb}

\end{document}